# Internal reverse-biased p-n junctions: a possible origin of the high resistance in phase change superlattice

Bowen Li[1], Longlong Xu[2], Yuzheng Guo[3], Huanglong Li[1,4*]

[1]Department of Precision Instrument, Center for Brain Inspired Computing Research, Tsinghua University, Beijing, China

[2]School of Materials Science and Engineering, Tsinghua University, Beijing, China

[3]College of Engineering, Swansea University, SA1 8EN, Swansea, UK

[4]Chinese Institute for Brain Research, Beijing, China

**\* Correspondence:**
Huanglong Li
li_huanglong@mail.tsinghua.edu.cn

**Abstract**

Phase change superlattice is one of the emerging material technologies for ultralow-power phase change memories. However, the resistance switching mechanism of phase change superlattice is still hotly debated. Early electrical measurements and recent materials characterizations have suggested that the Kooi phase is very likely to be the as-fabricated low-resistance state. Due to the difficulty in in-situ characterization at atomic resolution, the structure of the electrically switched superlattice in its high-resistance state is still unknown and mainly investigated by theoretical modellings. So far, there has been no simple model that can unify experimental results obtained from device-level electrical measurements and atomic-level materials characterizations. In this work, we carry out atomistic transport modellings of the phase change superlattice device and propose a simple mechanism accounting for its high resistance. The modeled high-resistance state is based on the interfacial phase changed superlattice that has previously been mistaken for the low-resistance state. This work advances the understanding of phase change superlattice for emerging memory applications.

# 1    Introduction

Resistive switching, the electrically induced convertible transition between a low resistance state (LRS) and a high resistance state (HRS), has been widely observed in two-terminal metal-insulator-metal (MIM) devices made from many classes of materials and has been the enabling phenomenon behind the workings of the emerging nonvolatile memory (NVM) devices for high-density data storage, in-memory computing and neuromorphic computing (Wang et al., 2020; Ielmini and Wong, 2018; Marković et al., 2020). Among various materials implementations and consequently different physical mechanisms, chalcogenide-based bulk crystalline-amorphous phase change (PC) is one of the most well studied. Discovered as long ago as the 1960s (Ovshinsky, 1968; Pearson, 1969), chalcogenide-based PC has gained rejuvenated interest for NVM since the 2000s (Raoux et al., 2008; Wuttig and Yamada, 2007) and has more recently given rise to the first commercially available NVM technology targeting mainstream computer systems (Hady et al., 2017).

Despite its relative maturity, PC-NVM devices suffer from a major bottleneck in further power reduction. The power issue arises due to the mass transport mechanism behind the bulk phase transition. To be specific, the bulk crystalline-amorphous PC requires, more or less, the rearrangements of all the atoms, which understandably costs sizable energy. This is especially severe for the amorphization process (RESET process from LRS to HRS) because it involves large atomic displacements in order to sufficiently randomize the atoms (melting the solid at high enough temperature to the liquid state and then quenching the melts).

Power reduction can be achieved through several routes, including device down-scaling, thermal confinement, and the exploitation of new materials and mechanisms (Wong et al., 2010). Among these optimization strategies, the use of the PC superlattice (SL) structure is very attractive and promising. PC-SL is a periodic structure of layers of two PC materials. It has initially been proposed to achieve multilevel storage because the mutually independent solid states of the two PC components increase the freedom of modulating the state of the device (Chong et al., 2006; Chong et al., 2008). SL composed of alternately stacked PC and non-PC confinement layers has also been used to improve the reliability of the PC operation (Shen et al., 2019; Ding et al., 2019). In 2011, Simpson et al. novelly proposed a new resistive switching mechanism in PC-SL, that is, confined atomic movements only at the interfaces between the two component PC materials (Simpson et al., 2011). This mechanism has been given the name "interfacial PC" (iPC). In contrast to bulk crystalline-amorphous PC, iPC is considered to be a crystalline-crystalline phase transition. In light of the origin of the high power consumption in the conventional PC process, the reduction of the number of atoms involved in the resistive switching process and the confinement of atomic displacements can in principle reduce the power consumption. Based on the experimental demonstrations of the several times' power reduction in the SET (HRS to LRS) and RESET operations, it was believed that iPC had occurred during the switching (Simpson et al., 2011). Very recently, further reduction of the RESET current has been achieved in PC-SL device with improved thermal management (Khan et al., 2021).

Since this iPC concept has been proposed, some experiments on PC-SL devices have been carried out by other groups, leading to hot debates on the resistive switching mechanism (Ohyanagi et al., 2013; Okabe et al., 2019; Ohyanagi and Takaura, 2016; Boniardi et al., 2019; Zhou et al., 2020). High-quality PC-SLs have also been fabricated and examined by high-resolution transmission electron microscopy. The experimental observations have indicated a different atomic structure (Momand et al., 2015; Lotnyk et al., 2017) from the previous intuitively predicted ones (Simpson et al., 2011; Tominaga et al., 2014; Ohyanagi et al., 2014). Therefore, the two early atomic-level iPC models (Simpson et al., 2011; Tominaga et al., 2014; Ohyanagi et al., 2014) have been modified (Kolobov et al., 2017; Chen et al., 2017).

The updated iPC model is based on the experimentally observed (Momand et al., 2015) Kooi structure (Kooi and De Hosson, 2002) of the PC-SL. In this Kooi structure, the Ge atoms are predominantly located in the centers of the nine-layer $Ge_2Sb_2Te_5$ blocks, adopting their preferred three-dimensional bonding environment, whereas the Sb atoms are predominantly located near the van der Waals (vdW) gaps between two blocks, adopting their preferred two-dimensional bonding environment. This is in contrast to the intuitively predicted SL structure composing individual GeTe and $Sb_2Te_3$ blocks (Simpson et al., 2011; Tominaga et al., 2014; Ohyanagi et al., 2014), despite the two have indeed been deposited alternately in respective steps. Accordingly, the newly proposed iPC mechanism states that the resistive switching is accompanied by the reconfiguration of the vdW gaps through the flipping of the SbTe bilayers near the vdW gaps and the subsequent bond reformation (Yu and Robertson, 2015), which results in non-stoichiometric $Ge_2SbTe_4$ and $Ge_2Sb_3Te_6$ blocks. These two non-stoichiometric component blocks can be regarded as being p-type and n-type doped. Due to local non-stoichiometry, the band gap originally present in the Kooi PC-SL (k-PC-SL) has vanished in the SbTe bilayer flipped PC-SL. Based on such band gap modulation, the former and latter structures have been interpreted as the HRS and LRS, respectively (Kolobov et al., 2017; Chen et al., 2017). Recently, a filamentary switching model has also been proposed, in which the k-PC-SL has also been interpreted as the HRS (Chen and Li, 2019).

However, these interpretations are not fully consistent with the experiment. Experimentally, the measured resistances of the HRS and LRS for iPC device were of the same order as those for normal PC device, respectively (Simpson et al., 2011; Li et al., 2018). It is also known that the cubic crystalline LRS of the normal PC device is metastable compared to the more stable hexagonal crystalline Kooi state, and the latter one has slightly lower resistance (Lankhorst et al., 2005). Therefore, the k-PC-SL must be the LRS rather than the HRS. To reconcile this contradiction, Saito et al. invoked the empirical findings of the p-type nature of the fabricated crystalline PC chalcogenides and considered the locally non-stoichiometric n-type $Ge_2Sb_3Te_6$ as compensating the natural p-type conductivity (Saito et al., 2019).

In this work, we argue that the SbTe bilayer flipped PC-SL can be more naturally interpreted as the HRS in iPC by viewing it as tandems of p-n junctions end to end. In this picture, no other empirical factor needs invoking to understand the origin of its high resistance. The LRS-HRS transition in PC-SL can be sufficiently understood as the result of the filed-driven reconfiguration of the vdW gap through SbTe bilayer flipping, as verified by our quantum transport simulations of the PC-SL model device.

## 2 HRS of PC-SL as tandems of p-n junctions end to end

Figure 1a and 1b illustrate the two-block supercell structures of the k-PC-SL and the SbTe bilayer flipped PC-SL, respectively. It should be noted that flipping only occurs at one of the two vdW gaps in the supercell structure (flipping at every two vdW gaps periodically, f-PC-SL), giving rise to periodic p-type $Ge_2SbTe_4$ and n-type $Ge_2Sb_3Te_6$ blocks. If flipping occurs at both vdW gaps (flipping at every vdW gap periodically, f*-PC-SL), the two new blocks will still be identical and stoichiometric but do not fully meet the two-dimensional and three-dimensional bonding preferences of Sb and Ge, respectively. A stability comparison between f-PC-SL and f*-PC-SL will be made in the next section. Here, we focus on f-PC-SL.

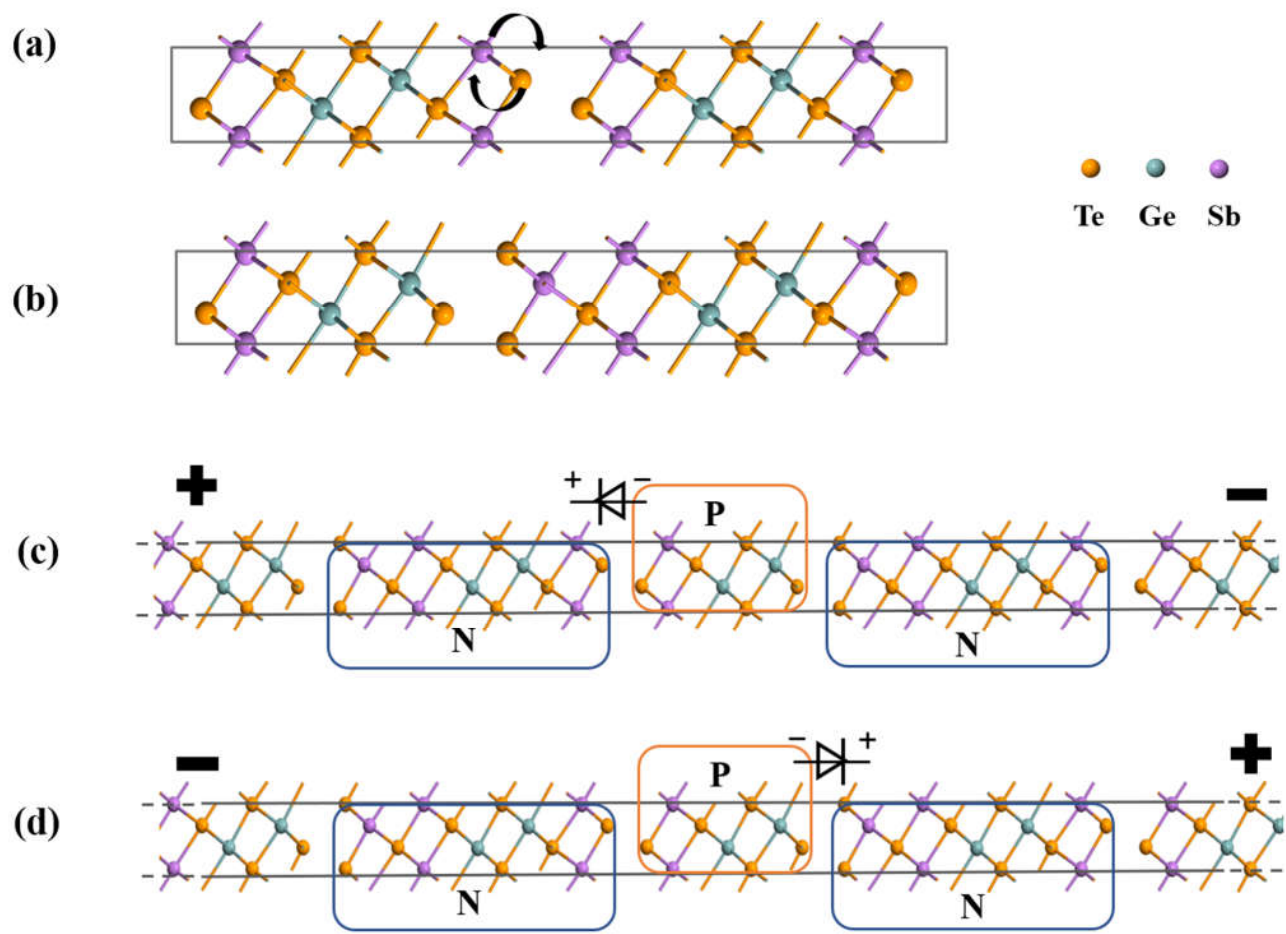

**Figure 1** (a) Two-block supercell structure of the k-PC-SL. Arrows indicate the flipping of the SbTe bilayer under electrical stimulus. (b) The structure of f-PC-SL. (c) and (d) The formation of internal reverse-biased p-n junction under voltages of different polarities.

It is clear that f-PC-SL could be viewed as tandems of $Ge_2SbTe_4$-$Ge_2Sb_3Te_6$ heterojunctions, or more generally, p-n junctions, end to end. As long as the experimentally deposited SL contains no less than three blocks, at least one formed p-n junction will be reverse-biased under either voltage polarity applied to the f-PC-SL, as sketched in figure 1c and 1d. It is this (these) reverse-biased p-n junction(s) that will contribute to the high resistance of the f-PC-SL.

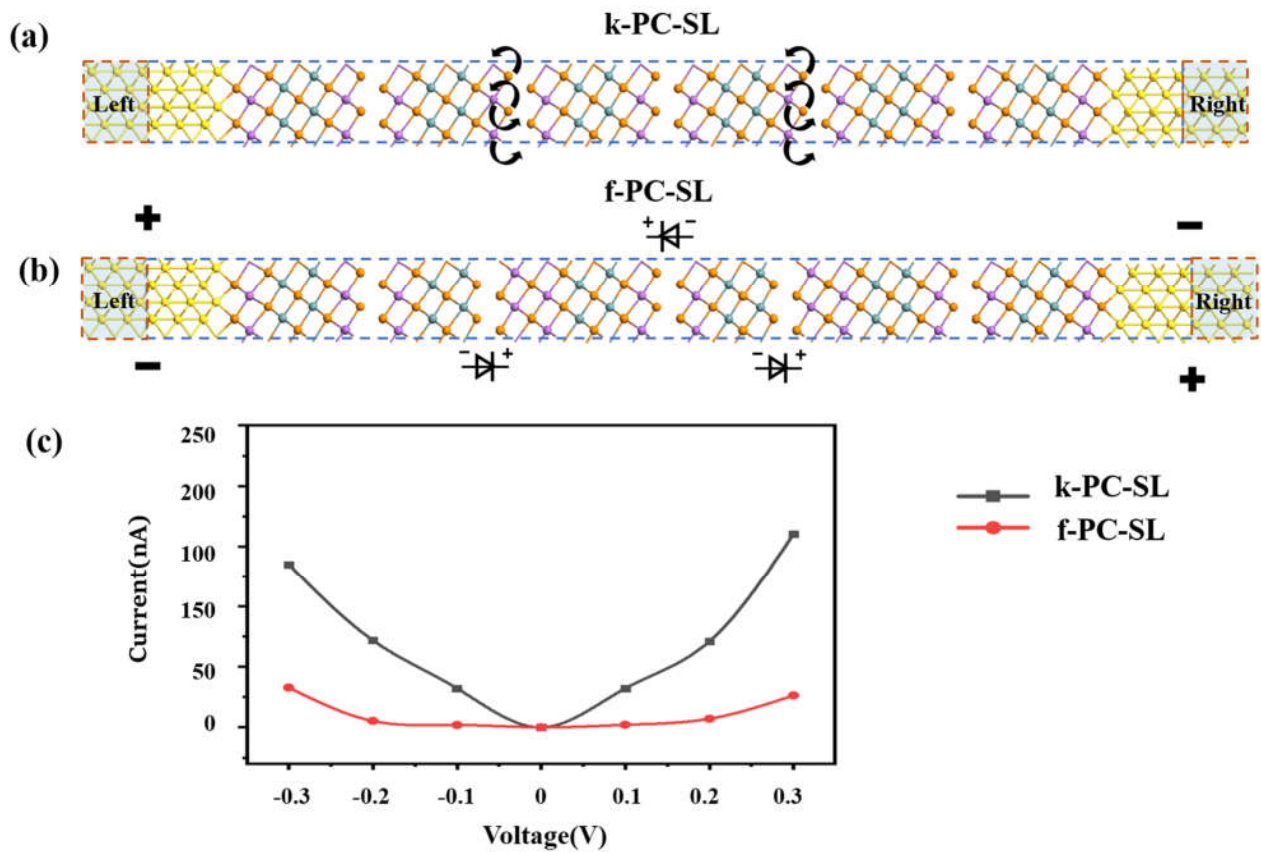

**Figure 2** (a) and (b) Atomic models of the k-PC-SL and f-PC-SL two-terminal devices, respectively. (c) I-V curves for the k-PC-SL and f-PC-SL devices.

To justify this argument, we carry out quantum transport simulations of two model devices, each composing a six-block PC-SL as the dielectric and a pair of symmetric Au electrodes, as shown in figure 2a and 2b for the k-PC-SL and f-PC-SL devices, respectively. In building the device models, Te layers are chosen to form direct contact with the Au electrodes considering that bonding between elements with large electronegativity difference is preferred. In this way, complete block structures can also be obtained. In addition, these finite-block PC-SL structures can be viewed as being cleaved from the bulk ones along the vdW gaps. This is reasonable because the inter-block vdW interactions are much weaker than the intra-block covalent interactions.

The simulated I-V curves for the k-PC-SL and f-PC-SL devices are shown in figure 2c. It is seen that the current values of the k-PC-SL device are always larger than those of the f-PC-SL device under the same voltages over the entire range. The currents in the f-PC-SL device are low because of the aforementioned reverse-biased p-n junction effect.

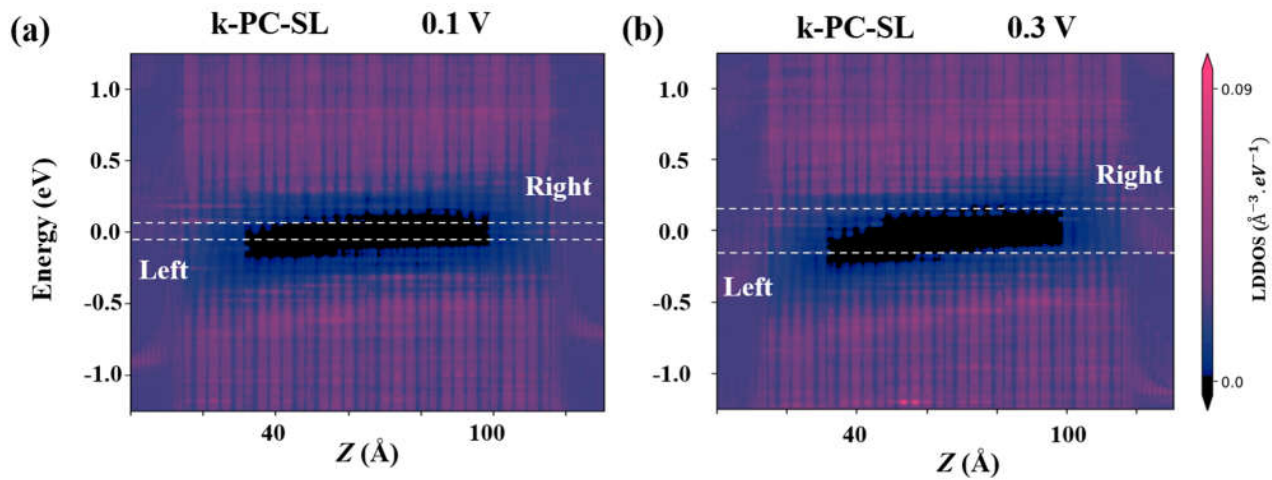

**Figure 3** (a) and (b) LDDOS for the k-PC-SL device under 0.1 V and 0.3 V voltage biases, respectively. The horizontal dashed lines indicates the Fermi levels of the left and right electrodes.

Figure 3 shows the local device density of states (LDDOS) for the k-PC-SL device under 0.1 V and 0.3 V voltage biases, respectively. It is seen that the contact resistance due to the formation of p-type Schottky barrier at the left Au: k-PC-SL interface is the main limiting factor for current conduction. As the voltage increases, the bands of k-PC-SL tilt more significantly and the thickness of the Schottky barrier decreases as a result of the band tilting. Both phenomena contribute to the increase of current.

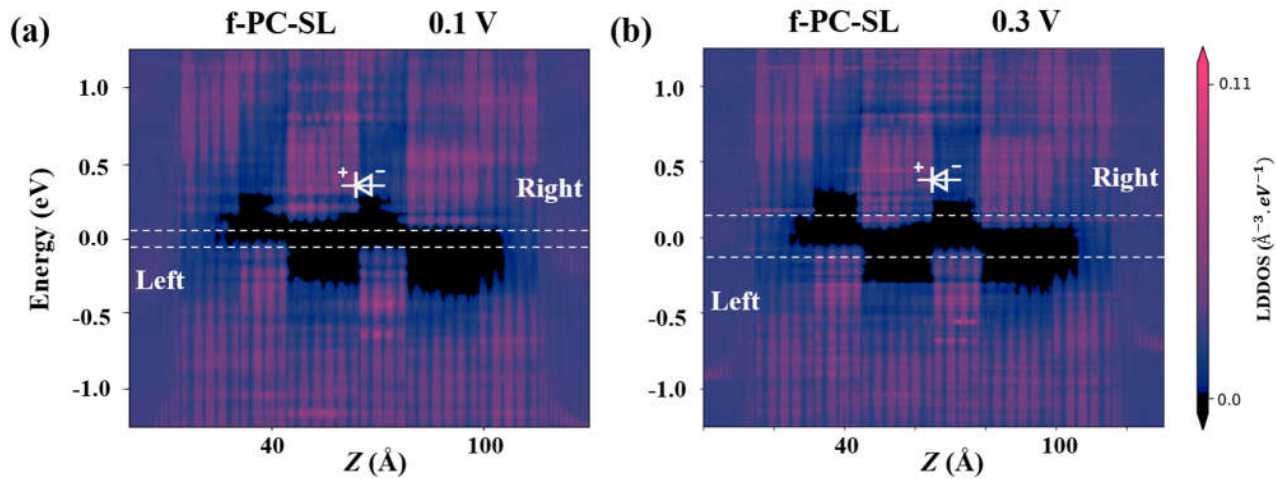

**Figure 4** (a) and (b) LDDOS for the f-PC-SL device under 0.1 V and 0.3 V voltage biases, respectively.

The band profile of the f-PC-SL device differs significantly from that of the k-PC-SL device. As shown in figure 4, the formation of tandems of p-n junctions results in undulated bands. Current conduction is mainly suppressed at the junctions between the p-type $Ge_2SbTe_4$ and n-type $Ge_2Sb_3Te_6$ blocks where the band line-ups are staggered. As the voltage increases, the energy barriers at the forward-biased p-n junctions are reduced. However, because of the co-existing reverse-biased p-n junctions, the total current is still very low.

The HRS origin of PC-SL due to internal reverse-biased p-n junction(s) may have important implication for multilevel operation of PC-SL device, as achieved recently (Khan et al., 2021). Ideally, each stoichiometric $Ge_2Sb_2Te_5$ block pair is independently switchable (see next section) to form an internal p-n junction. This scale-free behavior, reminiscent of that in ferroelectric $HfO_2$ with independently switchable polar layers (Lee et al., 2020), could be a basis for a multilevel PC-SL device whose number of states are similar to the number of the stoichiometric $Ge_2Sb_2Te_5$ block pairs.

## 3   Stability of the f-PC-SL structure

In the last section, we have compared the I-V characteristics and the LDDOS of the k-PC-SL device and f-PC-SL device, verifying the "internal reverse-biased p-n junction" assumption for understanding the higher resistance of the latter device. As introduced, this assumption is based on model structure containing tandems of p-type $Ge_2SbTe_4$ and n-type $Ge_2Sb_3Te_6$ blocks end to end which requires periodic SbTe bilayer flipping at every two vdW gaps, not at every vdW gap (f*-PC-SL). Figure 5 compares the free energies of f-PC-SL and f*-PC-SL over a wide temperature range. It is seen that the free energy of f-PC-SL is persistently lower than that of the f*-PC-SL, suggesting that the formation of f-PC-SL is more favorable and the model of tandems of p-n junctions end to end is valid.

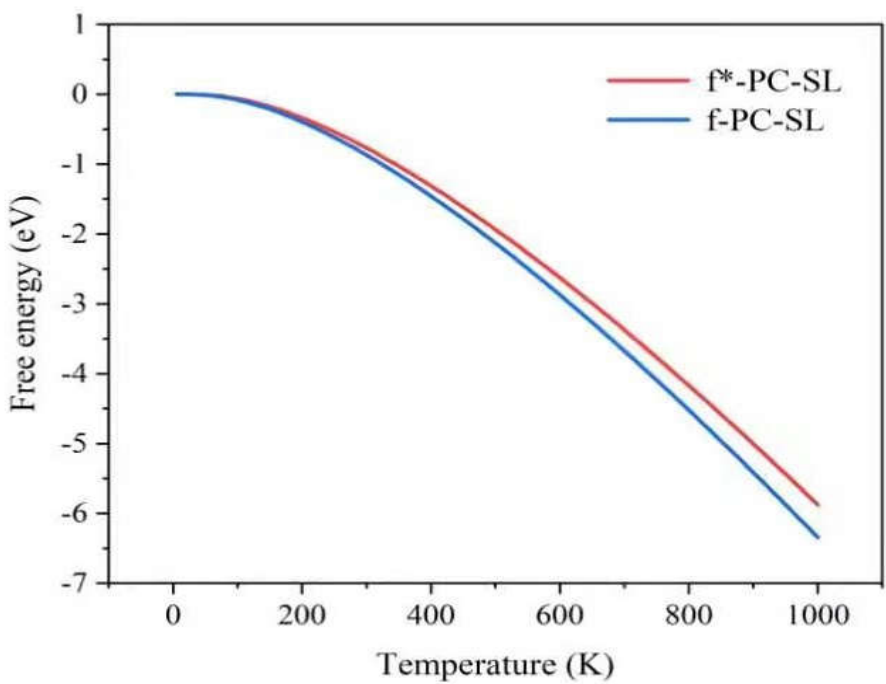

**Figure 5** Free energy against temperature for f-PC-SL and f*-PC-SL.

Finally, to justify f-PC-SL as a plausible high-resistance state, it is necessary to show that f-PC-SL is metastable and will not convert spontaneously back to k-PC-SL. To this end, we carry out transition state search to obtain the reaction energy diagram for f-PC-SL-to-k-PC-SL transition. As shown in figure 6, the two-block f-PC-SL supercell is calculated to be ~ 0.6 eV less stable than k-PC-SL. However, the energy barrier (or activation energy) for the occurrence of this transition is ~ 1.7 eV, which is sizable to guarantee the metastability of f-PC-SL.

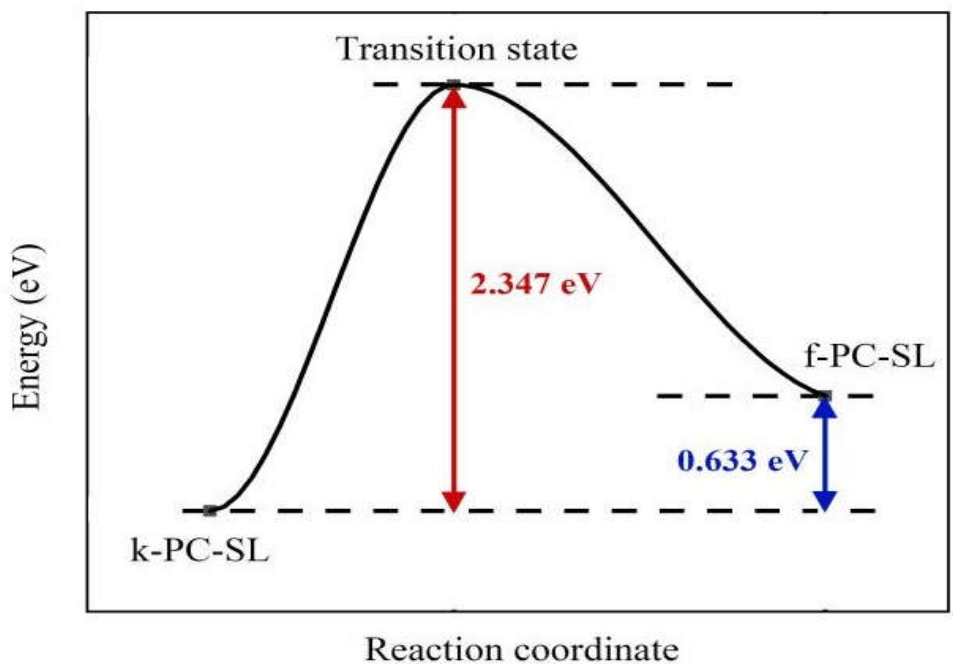

**Figure 6** Reaction energy diagram for f-PC-SL-to-k-PC-SL transition.

## 4    Conclusion

To conclude, we carry out atomistic transport modellings of the PC-SL devices and propose an internal reverse-biased p-n junction model to account for the high resistance of the interfacial phase changed device (f-PC-SL). We find that iPC via periodic SbTe bilayer flipping at very two vdW gaps effectively results in tandems of p-type and n-type blocks end to end. As long as the PC-SL contains no less than three blocks, at least one formed p-n junction will be reverse-biased under either voltage polarity applied to the PC-SL after iPC (f-PC-SL). It is this (these) reverse-biased p-n junction(s) that will contribute to the high resistance of the f-PC-SL. This work revises a mistake in the previous studies where f-PC-SL has been mistaken for the low-resistance state by total density of states analyses, and advances the understanding of PC-SL for emerging memory applications.

## 5    Methods

The PC-SL supercell models are simulated by CAmbridge Serial Total Energy Package (CASTEP) (Clark et al., 2005) based on density functional theory (DFT). The exchange-correlation potential is treated using the generalized gradient approximation (GGA) with the Perdew–Burke–Ernzerhof (PBE) flavour (Perdew et al., 1996). Van der Waals corrections based on DFT-D3 (Grimme et al., 2010) method are applied. Ultrasoft pseudopotentials are used with a plane wave cutoff energy of 380 eV. Quantum transport simulations are performed using the DFT-non-equilibrium Green's function (NEGF) approach (Brandbyge et al., 2002) by QuantumATK.

# 6 Conflict of Interest

The authors declare that the research was conducted in the absence of any commercial or financial relationships that could be construed as a potential conflict of interest.

# 7 Author Contributions

# 8 Funding

The authors acknowledge funding from National Natural Science Foundation (grant no. 61704096, 61974082 and 61836004). The authors acknowledge funding from National Key R&D Program of China (2018YFE0200200), Youth Elite Scientist Sponsorship (YESS) Program of China Association for Science and Technology (CAST) (no. 2019QNRC001), supercomputing wales project number scw1070, Tsinghua-IDG/McGovern Brain-X program, Beijing science and technology program (grant no. Z181100001518006 and Z191100007519009), the Suzhou-Tsinghua innovation leading program 2016SZ0102, and CETC Haikang Group-Brain Inspired Computing Joint Research Center.